\documentclass[3p,12pt]{elsarticle}
\usepackage{threeparttable,booktabs,tabularx}
\usepackage[fleqn]{amsmath}
\usepackage{cases}
\usepackage{multirow}
\usepackage{xcolor}
\usepackage[normalem]{ulem}
\usepackage{tabularx}
\usepackage{siunitx}
\usepackage{comment}
\usepackage{algorithm}
\usepackage{algpseudocode}
\usepackage{algorithmicx}
\usepackage{grffile}
\usepackage{rotating}
\usepackage{array}
\usepackage{lineno}
\usepackage{hyperref}
\usepackage{makecell}
\usepackage{graphicx,enumerate}
\usepackage{amssymb}
\usepackage{bm,amsmath}
\usepackage{subfigure}
\usepackage{caption,color}
\usepackage{threeparttable}
\usepackage{subeqnarray}
\usepackage{multirow}
\usepackage{lscape}
\usepackage{rotating}
\usepackage{graphics}
\floatname{algorithm}{Algorithm}
\usepackage{epstopdf}
\usepackage{adjustbox}
\journal{}
\usepackage{bm}

\biboptions{numbers,sort&compress} 

\begin{document}
	
\begin{frontmatter}

\title{Coherent Rayleigh-Brillouin scattering: influences of intermolecular potentials and chirp rates}

\author{Lei Wu\corref{mycorrespondingauthor1}}
\ead{wul@sustech.edu.cn}

\address{Department of Mechanics and Aerospace Engineering, Southern University of Science and Technology, 518055 Shenzhen, China}

\begin{abstract}
Chirped coherent Rayleigh-Brillouin scattering (CRBS) is a flow diagnostic technique that offers high signal-to-noise ratios and nanosecond temporal resolution. To extract information of dilute gas flow, experimental spectra must be compared with theoretical predictions derived from the Boltzmann equation. In this work, we develop a MATLAB code that deterministically solves the Boltzmann equation to compute CRBS spectra, enabling each line shape to be obtained in about one minute. We find that the CRBS spectrum is highly sensitive to the intermolecular potential, and that rapid chirping generates fine ripples around the Rayleigh peak along with spectral asymmetries.
\end{abstract}


\end{frontmatter}

\section{Introduction}\label{sec:1}

Rayleigh-Brillouin scattering is a non-intrusive flow diagnostic technique, which is categorized into spontaneous Rayleigh-Brillouin scattering (SRBS) and coherent Rayleigh-Brillouin scattering (CRBS). In SRBS, incident photons are scattered by spontaneous thermal density and pressure fluctuations within a medium, producing sidebands around the central Rayleigh peak, known as the Brillouin doublet \cite{Tenti1974}. The resulting spectral line shape reflects key thermodynamic and transport properties (such as viscosity, thermal conductivity, sound velocity, and flow velocity), making SRBS a powerful tool in atmospheric sensing and gas thermometry~\cite{srbs_scale,Witschas2014,Gu2014OL,GuRBS2015}, e.g., the Earth observation satellite ADM-Aeolus employs SRBS of ultraviolet laser light to directly measure global wind profiles from space. However, because the thermal fluctuation is spontaneous, the signal intensity is often weak.
In CRBS, light is scattered from the spatiotemporal density modulations induced by the interference of two laser pump beams. This externally driven excitation generates acoustic waves in the medium, which act as a dynamic diffraction grating for a probe beam. Owing to the use of phase-matching conditions, CRBS achieves a signal-to-noise ratio that is orders of magnitude higher than that of SRBS, enabling precise measurements even in low-density or weakly scattering environments~\cite{Pan2002,CRBS_JCP,Cornella,Shneider2013}. 

The CRBS spectrum depends on the frequency difference between the two pump lasers, or equivalently on the lattice velocity. To obtain the full line shape, the lattice velocity must be varied by stepwise tuning of one pump laser’s frequency. As a result, the diagnostic time can be on the order of minutes.
Recently, the use of chirped pump lasers has enabled CRBS measurements of gaseous flows with temporal resolutions on the order of 100 ns~\cite{barker2013,Suzuki2024PRA}. This technological advance opens the door to real-time, non-intrusive characterization of flow properties, even in rapidly evolving or transient environments. Such capability is particularly valuable for studying unsteady aerothermodynamic phenomena, shock-boundary layer interactions, and combustion processes, as well as for plasma diagnostics, where traditional methods often lack the required temporal resolution or risk perturbing the system under investigation.

To accurately extract gas information, the CRBS spectrum must be calculated with high precision. Because the scattering wavelength is comparable to the mean free path of gas molecules, the Boltzmann equation, rather than the Navier-Stokes equations, should be used.
However, since the Boltzmann equation is defined in a six-dimensional phase space and its collision operator involves a five-dimensional nonlinear integral, numerical solutions are computationally demanding. For instance, the spectrum has been obtained using the direct simulation Monte Carlo method~\cite{Bruno2019CPL,Suzuki2024PRA}, which essentially solves the Boltzmann equation in a stochastic manner~\cite{Bird1994}. Unfortunately, simulating CRBS with a chirped laser requires roughly 3 days for a single run, even when utilizing 32 processors on the Sherlock cluster at Stanford University~\cite{Suzuki2024PRA}.

So far, the prevailing method for calculating the RBS spectrum has been the Tenti method~\cite{Tenti1974}, in which the linearized Boltzmann collision operator is approximated using the first few eigenfunctions of Maxwellian molecules (where viscosity scales with temperature). As a result, each RBS line shape can be obtained within one second.
While this model yields good predictions for the SRBS spectrum, it performs less well for CRBS when the intermolecular potential deviates from the Maxwellian type~\cite{Wu-2022}, that is, the CRBS spectrum is sensitive to the intermolecular potential~\cite{Lei_AIP}. 

In addition to the influence of intermolecular potential, recent observation is that CRBS spectrum becomes asymmetric when a fast chirp rate is applied~\cite{Suzuki2024PRA}. However, because Monte Carlo simulations are highly time-consuming, the parameter space explored has been limited. In this paper, using the fast spectral method for solving the Boltzmann equation~\cite{Lei2013,LeiJFM2015}, we provide the Matlab code capable of calculating each line shape in about one minute, and systematically investigate how intermolecular potentials and chirp rates influence the CRBS line shape.


\section{The Boltzmann equation for monatomic gas}

The chirped pump lasers generate an optical lattice that exerts a dipole force on the gas molecules (assuming the optical lattice moves in the $x_2$ direction)~\cite{Suzuki2024PRA}
\begin{equation}\label{opt_lattice}
    F_2=\frac{I_0}{c\epsilon_0 \alpha k_L}\sin\left[k_Lx_2-\left(\omega_\text{min}t+\frac{\beta}{2}t^2\right)\right],
\end{equation}
where $I_0$ is the light intensity of the  pump beam, $c$ is the light speed, $\epsilon_0$ is the vacuum permittivity, $\alpha$ is the polarizability, $k_L$ is the wave number of the optical lattice, $t$ is the time; $\omega_\text{min}$ is the minimum laser frequency, and the chirp rate is
\begin{equation}\label{chirp_rate}
    \beta=\frac{\omega_\text{max}-\omega_\text{min}}{\tau},
\end{equation}
with $\omega_\text{max}$ being the maximum laser frequency and $\tau$ the time duration during which the laser is linearly chirped.

In gas kinetic theory, the velocity distribution function $f(t,x_2,\bm{v})$ is introduced to describe the state of a gas, where $\bm{v}=(v_1,v_2,v_3)$ the molecular velocity. 
Macroscopic quantities are obtained as moments of the distribution function, e.g., the number density $n$, bulk velocity $\bm{u}$, translational temperature $T_t$, pressure tensor ${p}_{ij}$, and translational heat flux $\bm{q}_t$ are given by
$\left[n, n\bm{u}, \tfrac{3}{2}nk_BT_t, p_{ij}, \bm{q}_t \right]
= \int \left[1, \bm{v}, \tfrac{m}{2}c^2, mc_ic_j, \tfrac{m}{2}c^2\bm{c}\right] f d\bm{v}$,
where $k_B$ is the Boltzmann constant, $m$ is the molecular mass, $\bm{c}=\bm{v}-\bm{u}$ is the thermal velocity, and the subscripts $i,j=1,2$, or 3.

The evolution of the distribution function is governed by the Boltzmann equation, where the terms on the left- and right-hand sides represent the streaming under the external acceleration and binary collisions of gas molecules, respectively:
\begin{equation}\label{Boltzmann}
\frac{\partial f}{\partial t}
+v_2\frac{\partial f}{\partial
	x_2}
    +\frac{F_2}{m} \frac{\partial f}{\partial v_2}
    =\iint
B(\theta,v_r)
[f(t,x_2,\bm{v}'_{\ast})f(t,x_2,\bm{v}')-
f(t,x_2,\bm{v}_{\ast})f]
d\bm{\Omega}
d\bm{v}_\ast,
\end{equation}
where $\bm v$ and $\bm v_\ast$ are the pre-collision molecular velocities, while $\bm v'$ and $\bm v'_{\ast}$ are  the post-collision velocities; conservation of momentum and energy yield the relations
$\bm{v}'=\bm{v}+(v_r\bm\Omega-\bm v_r)/{2}$ and $\bm{v}'_\ast=\bm{v}_\ast-(v_r\bm\Omega-\bm v_r)/{2}$,
with $\bm v_r=\bm v-\bm v_\ast$ the relative collision velocity and $\bm \Omega$ a unit vector pointing in the direction associated with the solid angle; the solid angle is the element of angular space that specifies the direction of the post-collision relative velocity. The deflection angle $\theta$ satisfies $\cos\theta=\bm \Omega\cdot \bm v_r/v_r$.

The collision kernel $B(\theta, v_r)$ is  determined by the intermolecular potential. It is the product of the differential cross-section and relative collision speed $v_r$, which is always positive.
Here, a modeled collision kernel for the inverse power-law potential is employed~\cite{Lei2013}: 
\begin{equation}\label{coll_kernel}
B(\theta, v_r)=B_0(\omega)\times
\sin^{\frac{1-2\omega}{2}}(\theta) 
\times{v}_r^{2(1-\omega)},
\end{equation}
where $\omega$ is the viscosity index, i.e., 
for a power-law potential, the viscosity scales with temperature as a power law with exponent $\omega$. In the cases of Maxwellian and hard-sphere gases, $\omega=1$ and 0.5, respectively. For other noble gases, the exponent satisfies $0.5<\omega<1$. For a Coulomb potential, $\omega=2.5$. However, due to Debye shielding, the effective viscosity index is reduced from this ideal value. 
Finally, $B_0(\omega)$ is a constant which will determine the gas viscosity. For example, for a hard-sphere gas of diameter $\sigma$, we have $B_0 = \sigma^2/4$, and according to the Chapman–Enskog expansion~\cite{CE}, the viscosity is
\begin{equation}
    \mu=1.016\frac{5\sqrt{mk_BT_t/\pi}}{16\sigma^2}.
\end{equation}


\subsection{The linearized Boltzmann equation}\label{linearization_FSM}

It is convenient to introduce the following dimensionless variables:
\begin{equation}\label{normalization}
\begin{aligned}
      \tilde{x}_2&=\frac{x_2}{\lambda}, \quad 
      (\tilde{\bm{v}}, \tilde{\bm{u}}, \tilde{\bm{c}})=\frac{(\bm{v},\bm{u},\bm{c})}{v_m}, \quad
 (\tilde{t},\tilde{\tau})=\frac{v_m}{\lambda}(t,\tau), \quad
 \tilde{a}_2=\frac{\lambda}{v_m^2}\frac{F_2}{m}, \\
 \tilde{\omega}_\text{min}&={\omega}_\text{min}\frac{\lambda}{v_m}, \quad 
\tilde{\beta}=\beta\frac{\lambda^2}{v^2_m}, \quad
\tilde{n}=\frac{n}{n_0},
\quad \tilde{f}=\frac{v_m^3}{n_0}f, 
\end{aligned}
\end{equation}
where $n_0$ is the average number density of gas molecules, $\lambda=2\pi/k_L$ is the scattering wave length, $v_m=\sqrt{2k_BT_0/m}$ is the most probable speed at the reference temperature $T_0$. For simplicity, the tilde will be removed.

Since the normalized acceleration in CRBS is very small, the distribution function can be expressed as
$f(t,x_2,\bm{v})=f_{eq}(\bm{v})+ 2({\pi{}I_0\alpha}/{c\epsilon_0k_BT_0}) h(t,x_2,\bm{v})$,
where $f_{eq}(\bm{v})=\pi^{-3/2}\exp(-v^{2})$ is the global  equilibrium distribution function. Furthermore, for the optical lattice~\eqref{opt_lattice}, a Fourier transform can be applied in the $x_2$ direction. As a result, the Boltzmann equation reduces to the following ordinary differential equation:
\begin{equation}\label{Chapter1_Boltzmann_lin}
\frac{\partial {h}}{\partial t}
+2\pi{i}v_2{h}
-\exp\left[i\left(\omega_\text{min}t+\frac{\beta}{2}t^2\right)\right]v_2f_{eq}
=\underbrace{\mathcal{L}^+(h) - \nu_{eq}(\bm{v}){h}}_{\mathcal{L}(h)},
\end{equation}
where $i$ is the imaginary unit, $
\mathcal{L}^+(h)=\iint B'  [f_{eq}(\bm{v}')h({\bm{v}}'_{\ast})+f_{eq}(\bm{v}'_\ast)h({\bm{v}}')-f_{eq}(\bm{v})h({\bm{v}}_\ast)]d\Omega d{\bm{v}}_\ast$ is the gain term of the linearized Boltzmann collision operator, 
with $B'=n_0\lambda{v^{1-2\omega}_m} B_0(\omega)\times
\sin^{\frac{1-2\omega}{2}}(\theta) 
\times{v}_r^{2(1-\omega)}$,
and 
\begin{equation}\label{collision_frequency}
    \nu_{eq}(\bm{v})=\iint{}B'f_{eq}(\bm{v}_{\ast}) d\Omega{d\bm{v}_\ast}
\end{equation}
is the equilibrium collision frequency.

The constant $B_0(\omega)$ is determined by the shear viscosity via the equation
$\mu(T_0)={mv_m} \int h_\mu(v)v_1v_2dv$,
where
$h_\mu(v)$ satisfies the integral equation $\mathcal{L}(h_\mu)=-2f_{eq}v_1v_2$. The fast spectral method and iterative scheme can be used to determine the constant $B_0(\omega)$~\cite{wuPoF2015}. This process is provided in the supplementary Matlab code.

\subsection{The linearized Shakhov kinetic model}

The Boltzmann collision operator is notoriously difficult to solve and is therefore often simplified through kinetic models, such as the Gross–Jackson model~\cite{GrossJackson1959}, the elliptic-statistic BGK model~\cite{Holway1966}, the Shakhov model~\cite{Shakhov_S}, and the Tenti S6 and S7 models~\cite{Tenti1974,Pan2005PRA}. In the linearized case, the collision operators in these kinetic models can be expressed as linear combinations of the eigenfunctions of the Boltzmann equation for a Maxwellian gas~\cite{wangchang_uhlenbeck_1952}, although the combination coefficients differ among models.
In our monograph~\cite{Wu-2022}, we showed that, for a monatomic gas, Tenti’s S6 model is equivalent to the linearized Shakhov model, while Tenti’s S7 model corresponds to the Gross–Jackson model. Although the S7 model incorporates stress terms in the collision operator, both experimental and numerical studies have demonstrated that the S6 model provides higher accuracy~\cite{Pan2004}.

In the linearized Shakhov model, the linearized Boltzmann collision operator $\mathcal{L}$ in Eq.~\eqref{Chapter1_Boltzmann_lin} is replaced by 
\begin{equation}
    \mathcal{L}_s=\frac{\sqrt{\pi}}{2\text{Kn}}\left\{ \left[\rho+2u_2v_2+T_t\left(v^2-\frac{3}{2}\right)+\frac{4q_{t2}v_2}{15}\left(c^2-\frac{5}{2}\right) \right]f_{eq} -h\right\},
\end{equation}
where Kn is the Knudsen number, defined as the ratio of the mean free path of gas molecules to the scattering wavelength:
\begin{equation}
  \text{Kn}=\sqrt{\frac{\pi}{4}} \frac{\mu(T_0){v_m}}{n_0k_BT_0\lambda}.  
\end{equation}
And the macroscopic quantities deviated from their corresponding values in equilibrium state, such as the number density $\rho$, bulk velocity $u_2$, temperature $T$, and translational heat flux $q_{t2}$ can be calculated as
\begin{equation}\label{MP}
[\rho,u_2,T_t,q_{t2}]=
\int\left[1,v_2,\frac{2}{3}\left(v^2-\frac{3}{2}\right), v_2\left({v^2}-\frac{5}{2}\right)\right]{h}d\bm{v},
\end{equation}
which, on top of the normalization~\eqref{normalization}, are further normalized by $2{\pi{}I_0\alpha}/{c\epsilon_0k_BT_0}$.

From the form of its collision operator, it is evident that the linearized Shakhov model does not capture the influence of the intermolecular potential. Running the supplementary Matlab code confirms that the Shakhov model produces line shapes similar to those obtained from the Boltzmann equation for a Maxwellian gas.


\section{CRBS spectra of monatomic gas}

Three factors affect the CRBS line shape. First, it is influenced by the Knudsen number, which is controlled by the shear viscosity.
Second, it is affected by the intermolecular potential, which is characterized by the viscosity index $\omega$ used in Eq.~\eqref{coll_kernel}. Third, the line shape depends on the generation of the optical lattice, for example, the chirp rate $\beta$ in Eq.~\eqref{opt_lattice}. 
In this section, these effects are investigated by solving the linearized Boltzmann equation deterministically, avoiding the computationally expensive stochastic direct simulation Monte Carlo method~\cite{Bruno2019CPL, Suzuki2024PRA}.

\subsection{Influence of intermolecular potential}

We first consider the case of a chirp-free optical lattice. To obtain the CRBS line shape for a chirp-free pulse, pump beams of different frequencies are superimposed and applied continuously throughout the simulation. In this case, the linearized Boltzmann equation~\eqref{Chapter1_Boltzmann_lin} is modified to include a new acceleration term (see the appendix in Ref.~\cite{LeiJFM2015}):
\begin{equation}
\frac{\partial {h}}{\partial t}
+2\pi{i}v_2{h}
-\frac{\sin(20\pi{t})}{t}v_2f_{eq}
=\mathcal{L}(h),
\end{equation}

The linearized Boltzmann collision operator $\mathcal{L}$ is efficiently solved using the fast spectral method~\cite{Lei2013,LeiJFM2015}, while the time-dependent ordinary differential equation is integrated by the second-order Heun’s scheme, with the initial condition $h=0$, a normalized time duration of $\tau = 30$, and time step $0.002$. During the simulation, the perturbed gas density $\rho(t)$ is recorded at each time step. After the simulation, a fast Fourier transform is applied to the time series of $\rho$, and the CRBS spectrum $S$ is obtained as the squared magnitude of the Fourier transform, i.e.,
\begin{equation}
    S=\left|\rho(t)\exp(-i\omega_a t)\right|^2,
\end{equation}
where $\omega_a$ is the angular frequency. 
For a given set of the Knudsen number and viscosity index, each line shape can be computed in about one minute using the in-house Matlab program $\text{CRBS}_{-}\text{monatomic}$ provided in the supplementary material.

\begin{figure}[t!]
	\centering
    {\includegraphics[width=0.6\textwidth,trim={0 0 30 30}, clip=]{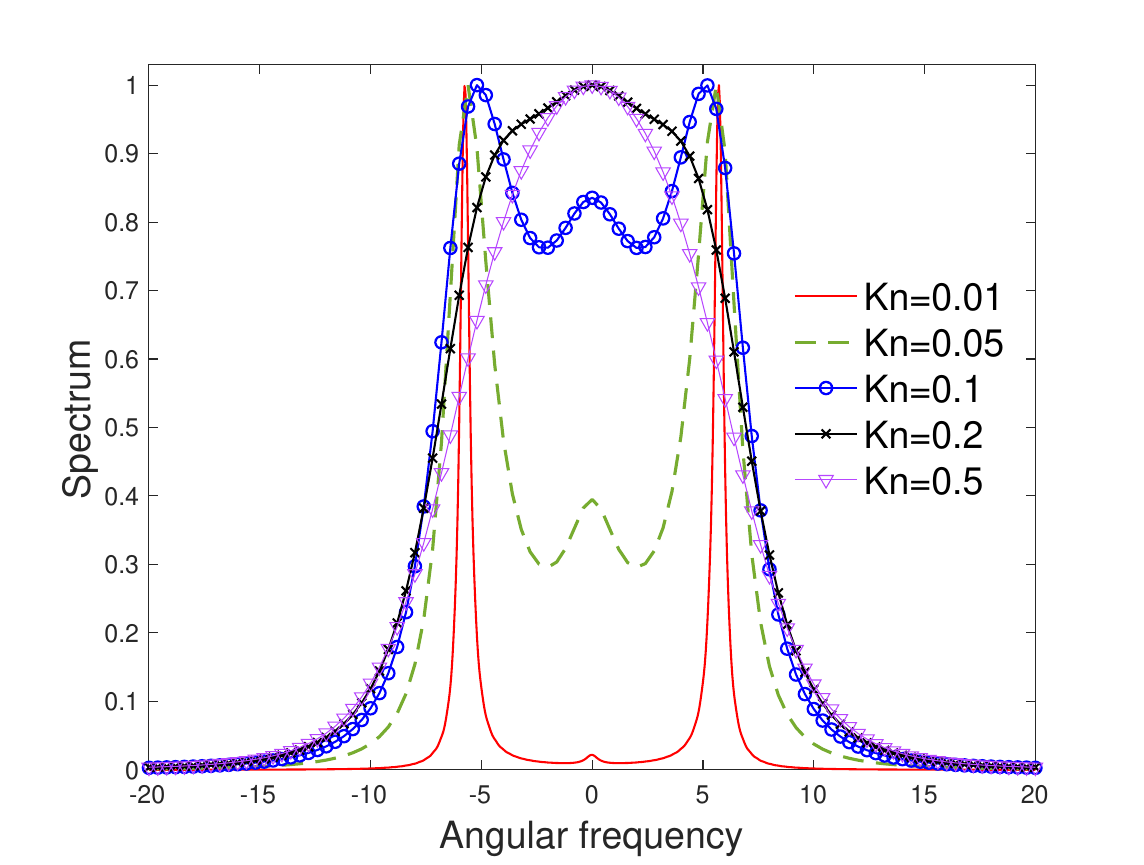}}
	\caption{Chirp-free CRBS spectra of a Maxwellian gas with $\omega = 1$ at different Knudsen numbers. In this and following figures, all spectra are normalized to their maximum magnitude.
	}
	\label{fig:chirp_free_Kn}
\end{figure}

Figure~\ref{fig:chirp_free_Kn} shows the typical CRBS spectra of Maxwellian gas at different Knudsen numbers. When Kn=0.01, the side Brillouin peaks dominate, while the central Rayleigh peak is almost negligible. The Brillouin peaks appear at the normalized angular frequency of $2\pi\sqrt{5/6}$, corresponding to an optical lattice velocity equal to the sound speed of a monatomic gas. This indicates that the Brillouin peaks arise from the propagation of sound waves in the gas. 
As the Knudsen number increases, kinetic effects become significant. The Brillouin peaks broaden, while the Rayleigh peak grows in magnitude. The enhancement of the Rayleigh peak is attributed to the suppression of collective sound-wave propagation at higher Knudsen numbers, where random thermal motions dominate over coherent acoustic oscillations. When Kn increases beyond about 0.15, the Rayleigh and Brillouin peaks start to merge, and the spectrum takes on a bell-shaped profile, dominated by stronger Rayleigh scattering from purely diffusive density fluctuations rather than from well-defined sound waves. When $\text{Kn} \gtrapprox 0.5$, the CRBS line shape is Gaussian: 
\begin{equation}\label{Gaussian_CRBS}
    S=\exp\left(-\frac{\omega_a^2}{4.84\pi^2}\right).
\end{equation}

\begin{figure}[t!]
	\centering
\subfigure[Kn=0.09]{\includegraphics[width=0.48\textwidth, trim={0 0 30 30}, clip]{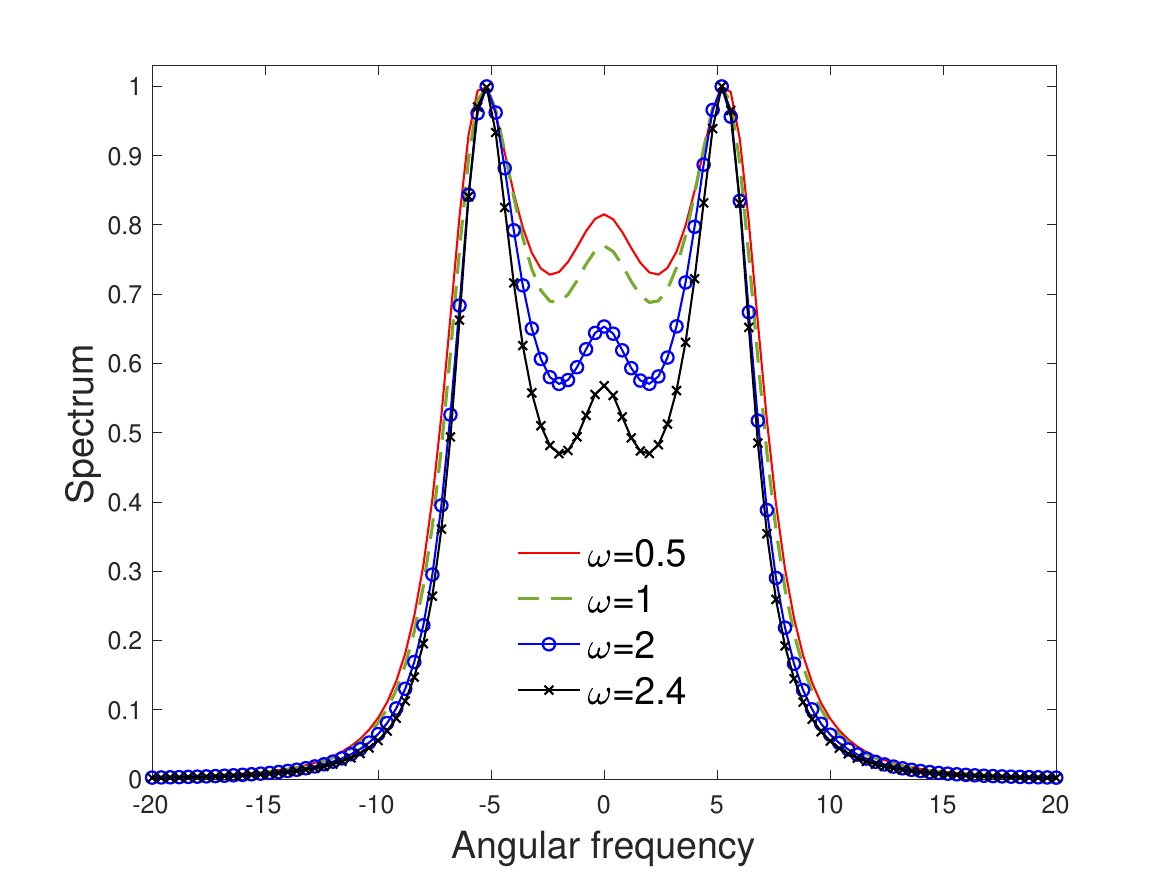}}
\subfigure[Kn=0.15]{\includegraphics[width=0.48\textwidth,trim={0 0 30 30}, clip=]{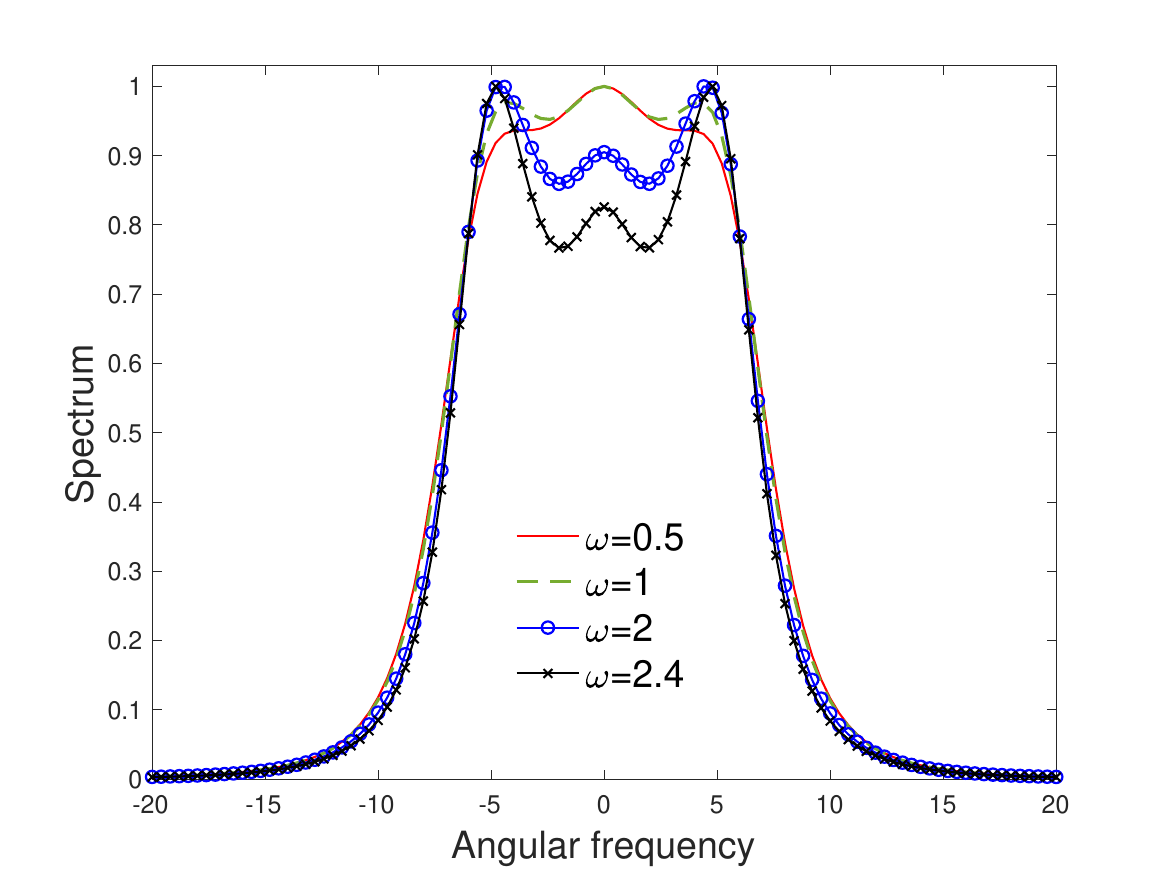}}
	\caption{
     Comparisons of chirp-free CRBS spectra for different power-law intermolecular potentials. 
	}
	\label{fig:chirp_free_omega}
\end{figure}

Now we analyze the role of intermolecular potential. 
When the Knudsen number is small, the gas dynamics is in the hydrodynamic regime, determined entirely by the Navier-Stokes equation and the transport coefficients (viscosity and thermal conductivity). In this case, the intermolecular potential has no influence. Similarly, when the Knudsen number is large and the gas dynamics enters the free-molecular regime, the Boltzmann collision operator $\mathcal{L}$ vanishes, hence the influence of the intermolecular potential is absent, and the spectrum becomes Gaussian. 
Thus, the intermolecular potential only plays a role in the kinetic regime, see Fig.~\ref{fig:chirp_free_omega}. For example, when Kn = 0.09, as the viscosity index increases from 0.5 (hard-sphere gas) to 2.4 (representing a soft potential close to the Coulomb potential\footnote{Note that the Boltzmann equation cannot be directly applied for the Coulomb potential since the viscosity cross-section diverges. One possible remedy is to introduce a cutoff in the deflection angle, or alternatively to employ the Fokker–Planck approximation in the grazing-collision limit. These approaches, however, lie beyond the scope of the present study.}), the magnitude of the Rayleigh peak decreases. Moreover, the closer the viscosity index is to that of the Coulomb potential, the more rapidly the central spectrum decreases.

This phenomenon can be explained in terms of the collision frequency defined in Eq.~\eqref{collision_frequency}. For Maxwellian molecules ($\omega = 1$), the collision kernel~\eqref{coll_kernel} is independent of the relative collision speed, and consequently the equilibrium collision frequency~\eqref{collision_frequency} does not depend on the molecular velocity. In contrast, for a hard-sphere gas ($\omega = 0.5$), the equilibrium collision frequency increases with the molecular speed $v$, see Fig.~13 in Ref.~\cite{Lei2013}. This implies that for $v \approx 0$, the collision frequency is lower than that of a Maxwellian gas, corresponding to an effectively larger Knudsen number (fewer collisions, larger viscosity, and hence a larger Knudsen number). Since at $\text{Kn} \approx 0.09$ the magnitude of the Rayleigh peak increases with the Knudsen number (see Fig.~\ref{fig:chirp_free_Kn}), the Rayleigh peak of the hard-sphere gas is higher than that of the Maxwellian gas, as shown in Fig.~\ref{fig:chirp_free_omega}(a).
By contrast, for a soft potential with $\omega > 1$, the equilibrium collision frequency decreases with increasing molecular speed $v$. Consequently, for $v \sim 0$, the effective Knudsen number is smaller, and the magnitude of Rayleigh peak is reduced.

\subsection{Influence of chirp rate}

\begin{figure}[t!]
	\centering
\subfigure[Kn=0.01]{\includegraphics[width=0.48\textwidth, trim={0 0 30 30}, clip]{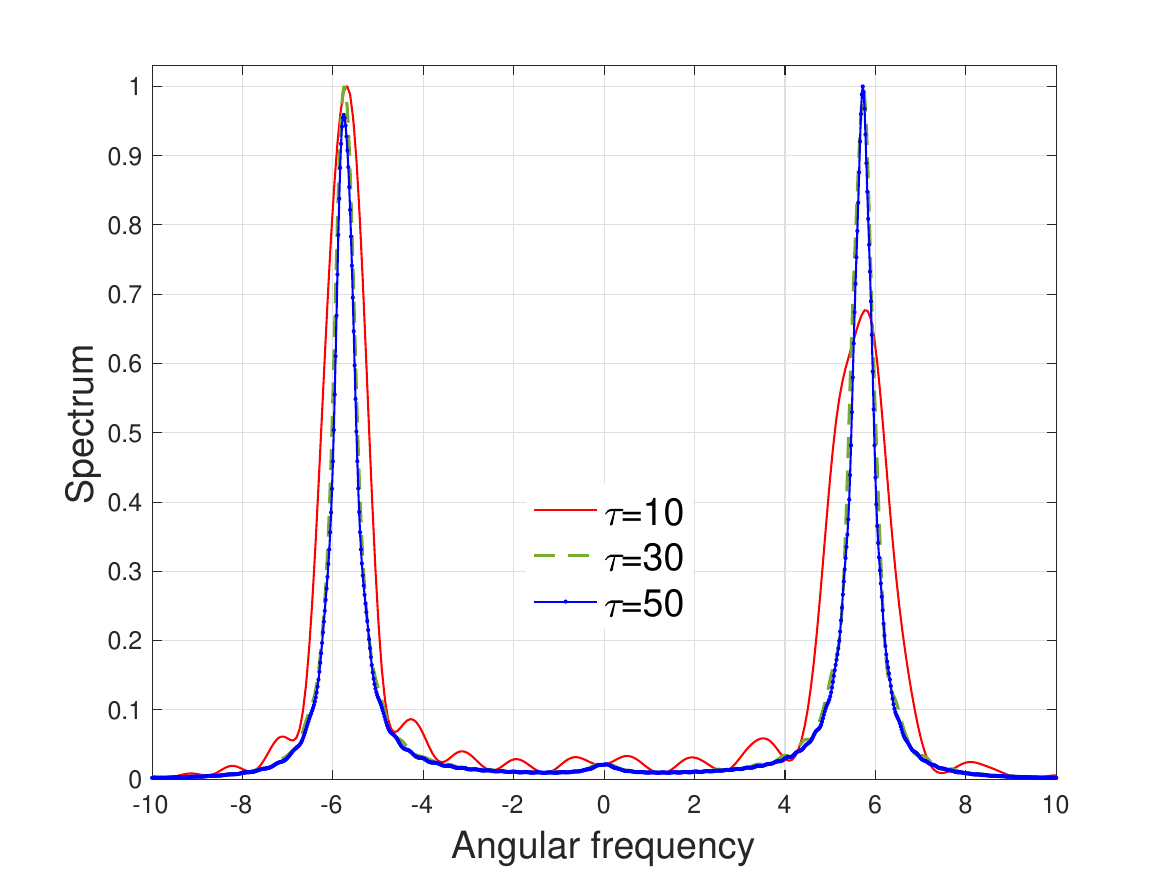}}
\subfigure[Kn=0.09]{\includegraphics[width=0.48\textwidth, trim={0 0 30 30}, clip]{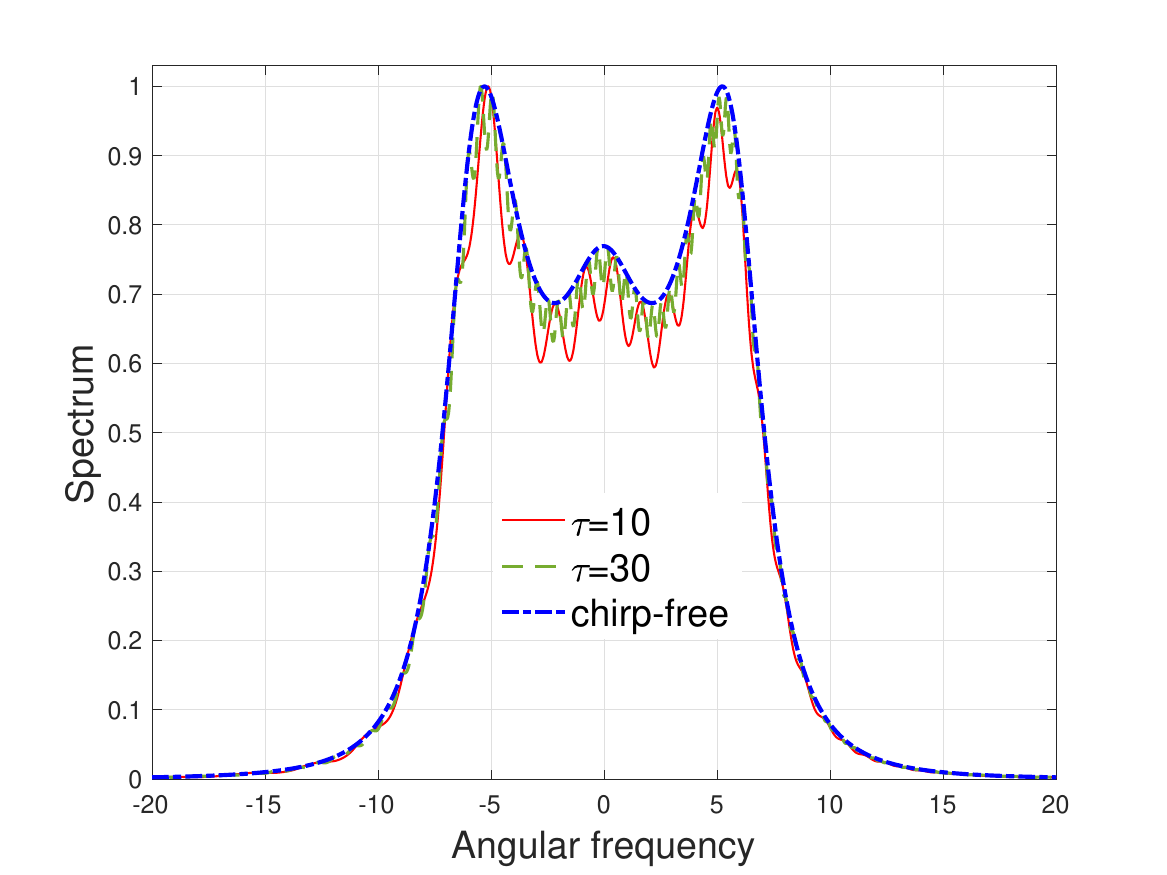}}\\
\subfigure[Kn=0.12]{\includegraphics[width=0.48\textwidth, trim={0 0 30 30}, clip]{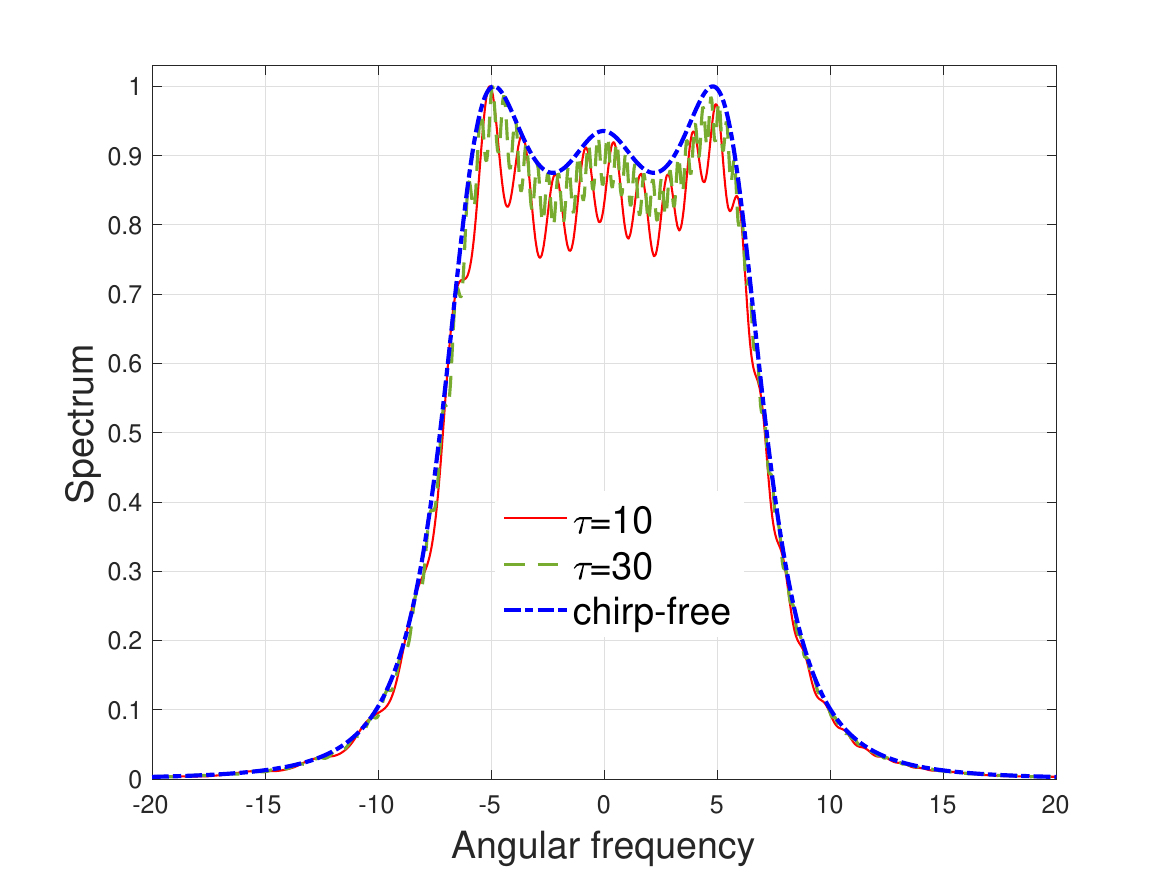}}
\subfigure[Kn=0.15]{\includegraphics[width=0.48\textwidth, trim={0 0 30 30}, clip]{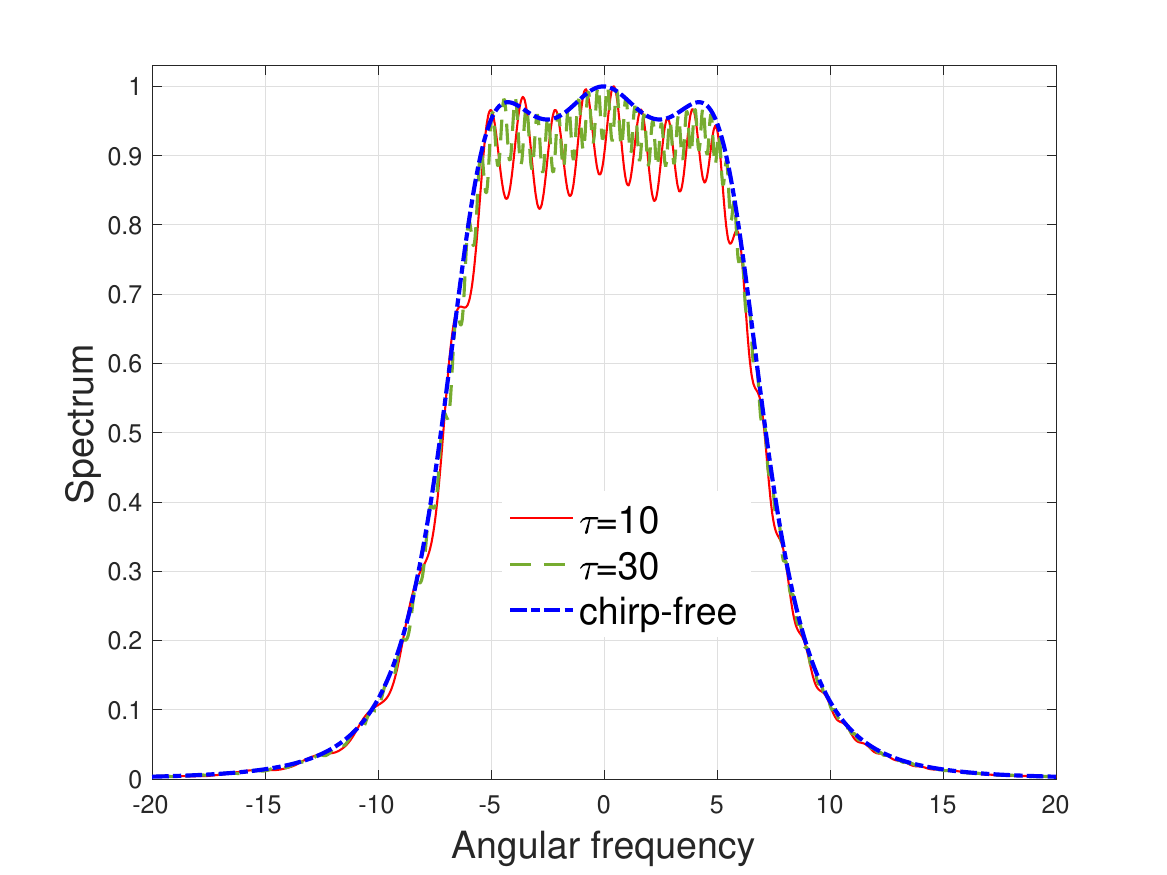}}
	\caption{
     Chirped CRBS spectra for Maxwell gas at different chirp duration $\tau$. 
	}
	\label{fig:chirp_Maxwell}
\end{figure}

Now, we solve the linearized Boltzmann equation~\eqref{Chapter1_Boltzmann_lin} with the chirped optical lattice to investigate the effect of the chirp rate $\beta$. We choose the normalized angular frequency range  $\omega_\text{min}=-8\pi$ and $\omega_\text{max}=8\pi$, corresponding to lattice speeds in the range $[-4v_m,4v_m]$, which is sufficiently wide to cover the entire line shape. The chirp rate is controlled by the chirp duration $\tau$ in Eq.~\eqref{chirp_rate}, which is normalized by $\lambda/v_m$. The Maxwellian gas is considered. 

Figure~\ref{fig:chirp_Maxwell}(a) shows the CRBS spectra at three different chirp durations for Kn = 0.01. At a fast chirp rate ($\tau = 10$), the line shape becomes clearly asymmetric: the right Brillouin peak is significantly lower than the left one, consistent with the observations reported in Ref.~\cite{Suzuki2024PRA}. Furthermore, we observe that the central Rayleigh peak develops fine ripples. When $\tau = 30$, the CRBS spectrum regains symmetry and these ripples nearly vanish. At $\tau = 50$, however, a slight asymmetry reemerges, but now the right Brillouin peak becoming slightly higher than the left.

Figure~\ref{fig:chirp_Maxwell}(b) shows the CRBS spectra at $\text{Kn} = 0.09$. For the chirped laser, slight asymmetries are observed, with the right Brillouin peak appearing lower than the left. In addition, fine ripples emerge in the Rayleigh peak. As the Knudsen number increases further, the Rayleigh and Brillouin peaks begin to overlap, leading to a flattened line shape near the central frequency, as shown in Fig.~\ref{fig:chirp_Maxwell}(c, d). In this regime, the spectral asymmetry becomes less discernible, although the fine ripples persist. These results indicate that the pronounced spectral asymmetry originates from collisional effects of the gas molecules.

In all cases, the oscillation amplitude of the fine ripples gradually decreases as the chirp duration increases (i.e., as the chirp rate decreases). However, the oscillation frequency tends to increase with the chirp duration. In fact, Fig.~\ref{fig:chirp_Maxwell} indicates that the oscillation period (i.e., the number of ripples) is approximately equal to the normalized chirp duration $\tau$, independent of the Knudsen number. Thus, this behavior cannot be attributed to collisional effects of the gas molecules. At sufficiently large values of $\tau$, the spectrum eventually converges to the chirp-free case described in Fig.~\ref{fig:chirp_free_omega}(a).


\section{CRBS of polyatomic gas}

The case of a polyatomic gas is considerably more complex than that of a monatomic gas. This is because the governing kinetic equation—the Wang-Chang and Uhlenbeck equation~\cite{WangCS}, which generalizes the Boltzmann equation to account for quantum energy levels associated with rotational and vibrational motions—is significantly more difficult to solve. This is why the simplified Tenti model~\cite{Tenti1974,Pan2005PRA} is commonly employed to calculate the SRBS and CRBS spectra. Clearly, this model does not account for the general intermolecular potential, but only for Maxwellian potential where $\omega=1$. 

In this paper, we employ the kinetic model for non-vibrating polyatomic gases~\cite{LeiJFM2015}, in which translational motion is described by the Boltzmann collision operator, allowing the effects of the intermolecular potential to be captured. Only internal rotational motion is considered, since the scattering frequency in CRBS is sufficiently large that vibrational motion remains effectively frozen. 
In this case, in addition to the Knudsen number, viscosity index, and chirp rate, the CRBS spectrum also depends on the rotational degrees of freedom $d_r$, the rotational collision number $Z$, the translational Eucken factor $f_{t}$, and the rotational Eucken factor $f_{r}$.

The rotational collision number $Z$ characterizes the rate of energy exchange between rotational and translational modes, as described by the Landau–Teller equation:
\begin{equation}
    \frac{d{T_t}}{d t}=-\frac{d_r}{3}\frac{T_t-T_r}{Z\mu(T_0)/n_0k_BT_0},
\end{equation}
where $T_r$ denotes the rotational temperature. Because thermal equilibrium with translational motion is reached only after a finite relaxation time, a polyatomic gas exhibits a bulk viscosity given by ${\mu_b}=[{2d_r}/{3(d_r+3)}]Z\mu$.

The study of molecular thermal conductivity is more complex than that of bulk viscosity, particularly in the context of Rayleigh–Brillouin scattering applications.
The heat flux of a molecular gas consists of two contributions: one from the transfer of kinetic energy due to molecular translational motion, and the other from the transfer of internal energy due to molecular diffusion. Historically, Eucken first expressed the relationship between the molecular thermal conductivity $\kappa$ and the molecular shear viscosity $\mu$ as
\begin{equation}\label{Enckenfactors}
\begin{aligned}
\frac{\kappa m}{\mu k_B}=&\frac{\kappa_{t} m}{\mu k_B}+\frac{\kappa_{r} m}{\mu k_B}
+\frac{\kappa_{v} m}{\mu k_B}\\
&\equiv\frac{3}{2}f_{t}+\frac{d_r}{2}f_{r}+\frac{d_v}{2}f_{v},
\end{aligned}
\end{equation}
where $d_v$ is the vibrational degrees of freedom, $\kappa_{t}$, $\kappa_{r}$, and $\kappa_{v}$ are the translational, rotational and vibrational components of the thermal conductivity of gas molecules, respectively, while $f_{t}$, $f_{r}$, and $f_v$ are the translational, rotational, and vibrational Eucken factors, respectively. For monatomic gases, $f_{t}$ is very close to $5/2$. However, for molecular gases, the values of $f_{t}$, $f_{r}$, and $f_{v}$ are difficult to determine, since experiments measure only the total heat flux rather than its individual components.

Here, the following kinetic model proposed by Wu \textit{et. al.} is employed~\cite{LeiJFM2015,Wu2020JFM}:
\begin{equation}\label{SRBS_poly}
\begin{aligned}
\frac{\partial{h}}{\partial {t}}&+2\pi{i}v_2{h}
-\exp\left[i\left(\omega_\text{min}t+\frac{\beta}{2}t^2\right)\right]v_2f_{eq}=\mathcal{L}(h)\\
& 
+\sqrt{\frac{\pi}{4}}\frac{1}{\text{Kn}}\left[\frac{T-T_t}{Z}\left(v^2-\frac{3}{2}\right)
+\left(\frac{8}{15}-\frac{4}{3f_t}\right)q_{t2}v_2\left(v^2-\frac{5}{2}\right)\right]{f_{eq}}, \\
\frac{\partial{h_2}}{\partial {t}}&+2\pi{i}v_2{h_2}
-\left(1-\frac{d_r}{2}\right)\exp\left[i\left(\omega_\text{min}t+\frac{\beta}{2}t^2\right)\right]v_2f_{eq}
=\sqrt{\frac{\pi}{4}}\frac{1}{\text{Kn}}\left[\frac{d_r}{2}T_rf_{eq}-h_2\right]\\
&
+ \frac{d_r}{2Z}\sqrt{\frac{\pi}{4}}\frac{1}{\text{Kn}}(T-T_r)f_{eq}
+2\sqrt{\frac{\pi}{4}}\frac{1}{\text{Kn}}\frac{f_r-1}{f_r} q_{r2}v_2f_{eq},
\end{aligned}
\end{equation}
where $T=(3T_t+d_rT_r)/(3+d_r)$, the rotational temperature is $T_r=(2/d_r)\int{h_2}d\bm{v}$, and the rotational heat flux is $q_{r2}=\int{v_2h_2}d\bm{v}$. For a given set of the Knudsen number and viscosity index, each line shape can be computed in about one minute using the in-house Matlab program $\text{CRBS}_{-}\text{polyatomic}$ provided in the supplementary material.

\begin{figure}[t!]
	\centering
    \subfigure[Maxwellian gas, Kn=0.04]{\includegraphics[width=0.48\textwidth, trim={0 0 30 30}, clip]{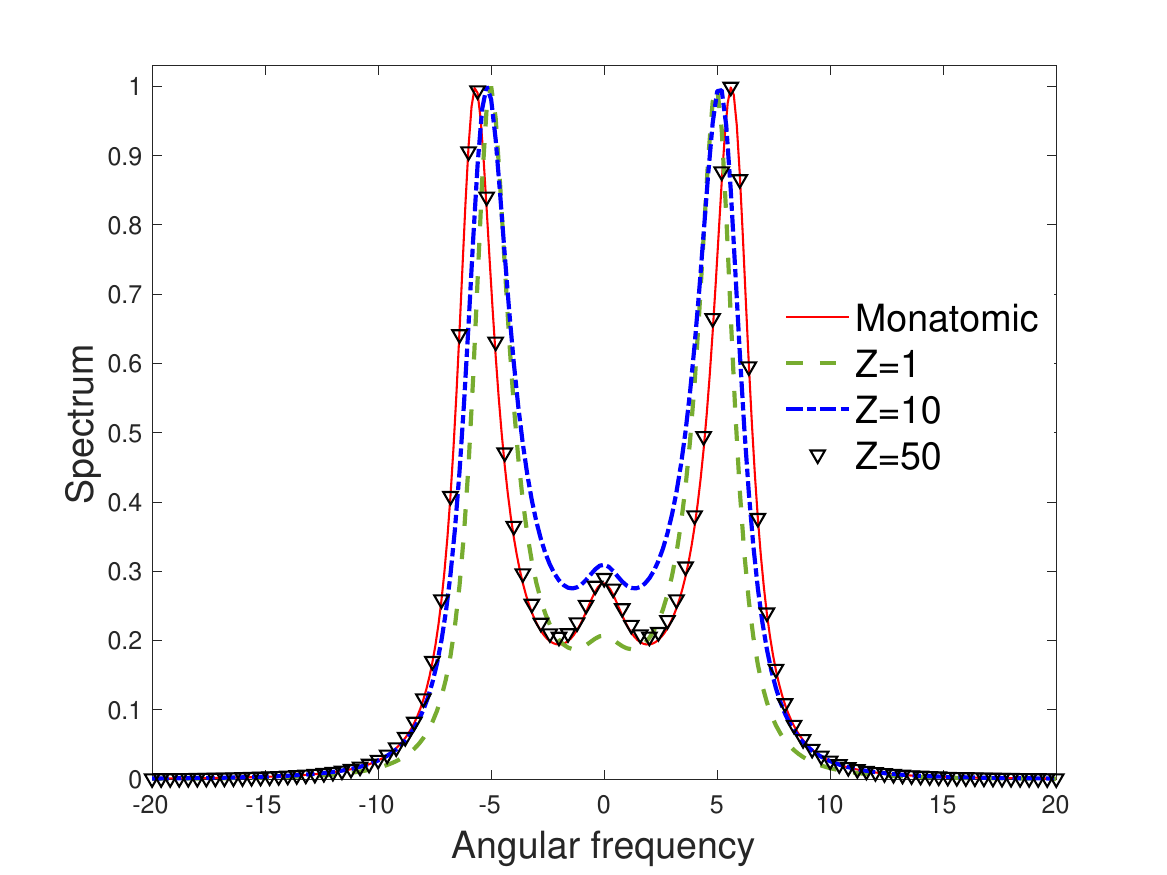}}
\subfigure[soft potential with $\omega=2.4$, Kn=0.04]{\includegraphics[width=0.48\textwidth, trim={0 0 30 30}, clip]{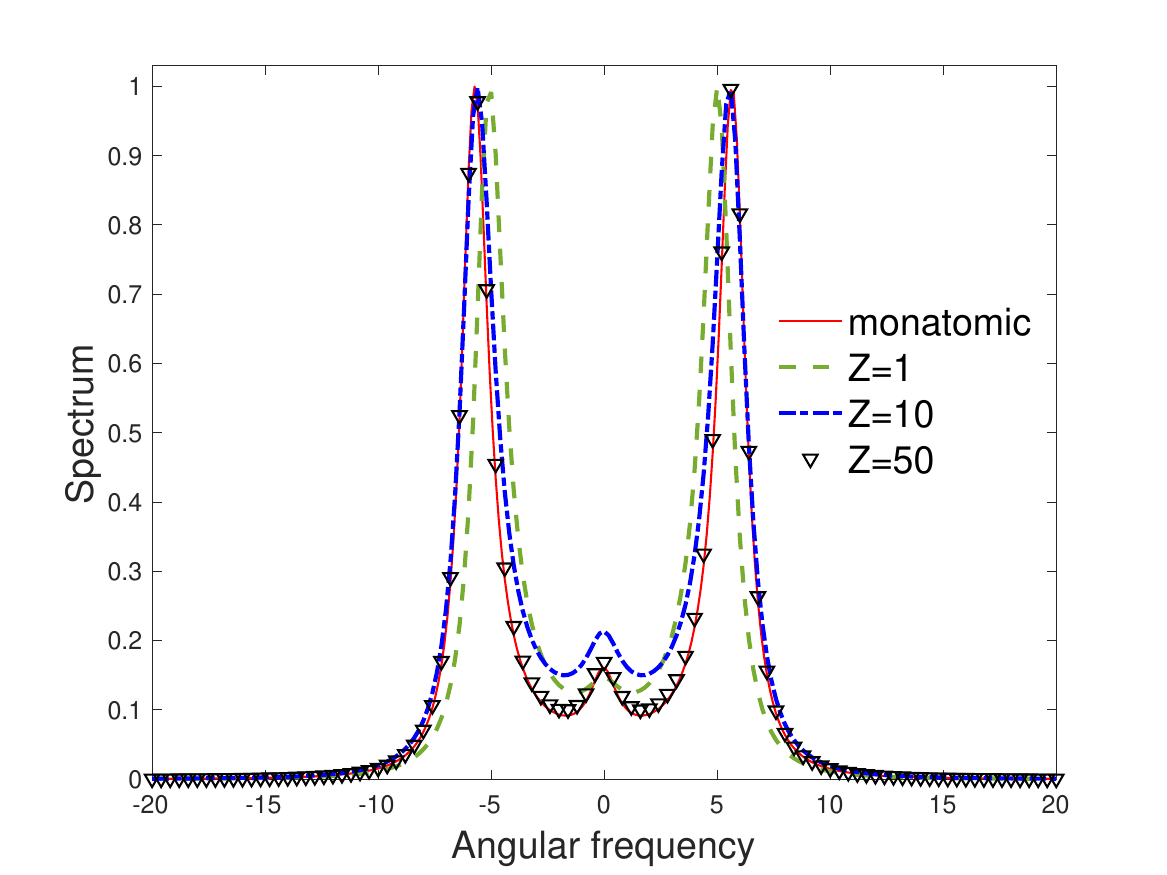}}
\subfigure[Maxwellian gas, Kn=0.09]{\includegraphics[width=0.48\textwidth, trim={0 0 30 30}, clip]{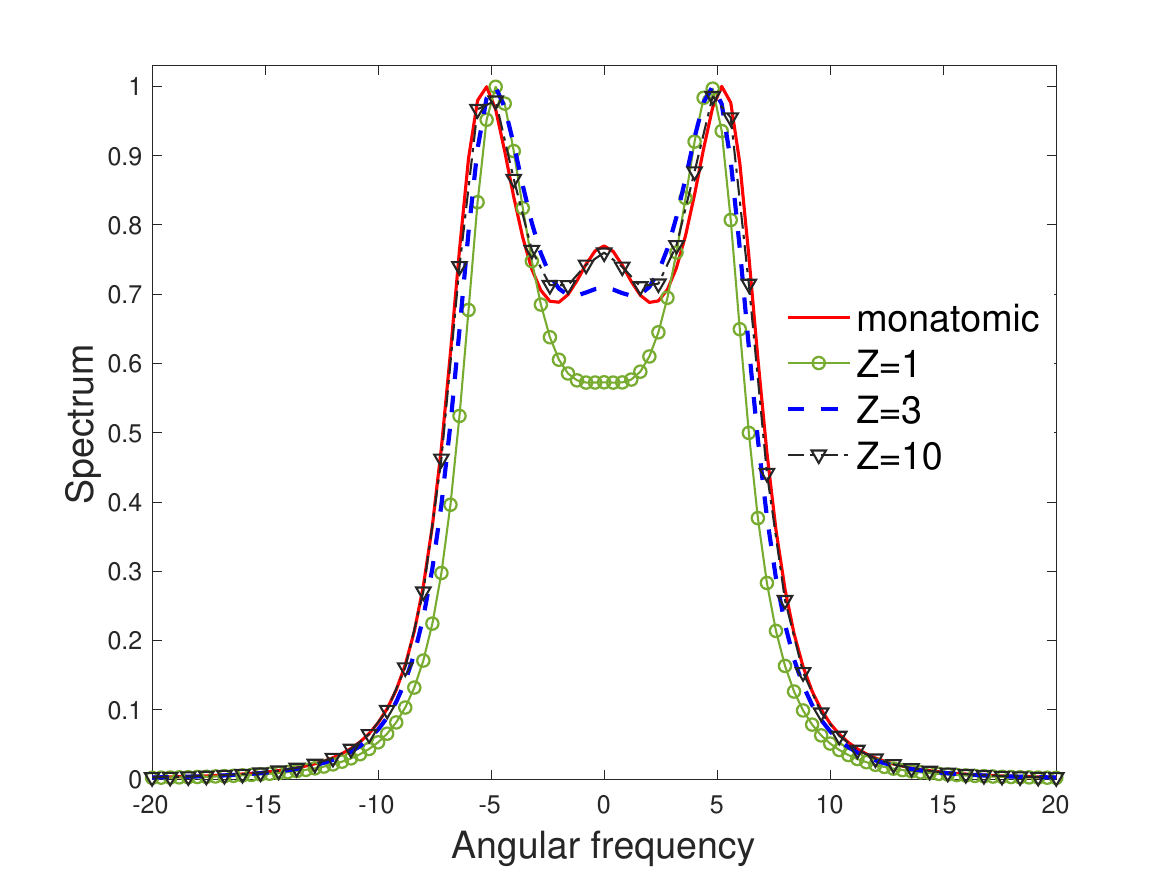}}
\subfigure[soft potential with $\omega=2.4$, Kn=0.09]{\includegraphics[width=0.48\textwidth, trim={0 0 30 30}, clip]{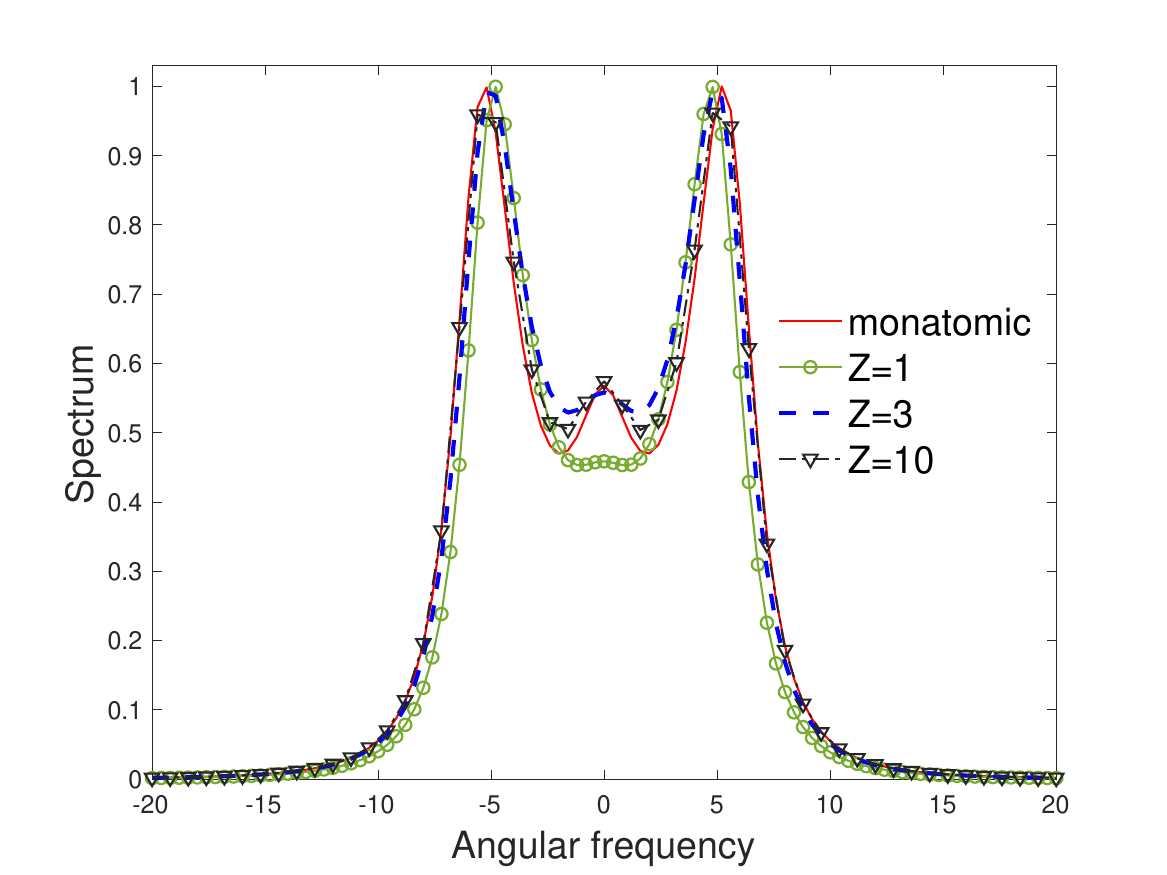}}
	\caption{
     Chirp-free CRBS spectra of polyatomic gas with different values of rotational collision number $Z$, with rotational degrees of freedom $d_r=2$, translational Eucken factor $f_t=2.4$, and rotational Eucken factor $f_r=1.5$. 
	}
	\label{fig:chirp_free_polyatomic_dr2}
\end{figure}

Figure~\ref{fig:chirp_free_polyatomic_dr2} presents typical chirp-free CRBS spectra of a polyatomic gas with two rotational degrees of freedom, shown for both a Maxwellian gas and a soft potential with $\omega = 2.4$. For comparison, the corresponding spectra of a monatomic gas are also included. 
At Kn = 0.04, when the rotational collision number $Z$ is small, the rotational degrees of freedom are efficiently excited, leading to Brillouin peaks at an angular frequency of $2\pi\sqrt{7/10}$, which corresponds to the sound speed of a polyatomic gas with $d_r = 2$. In contrast, when $Z = 50$, the exchange between translational and rotational energies becomes slow, and the rotational motion can effectively be regarded as frozen. In this limit, the dynamics of the polyatomic gas reduce to those of a monatomic gas, and the Brillouin peak shifts to $2\pi\sqrt{5/6}$.
At Kn = 0.09, a small value of $Z$ reduces the magnitude of the Rayleigh peaks, and the line shape of the polyatomic gas rapidly approaches that of a monatomic gas at smaller $Z$. This occurs because translational collisions are weaker than in the case of Kn = 0.04, causing the rotational motion to freeze earlier.

\begin{figure}[t!]
	\centering
{\includegraphics[width=0.6\textwidth, trim={0 0 30 30}, clip]{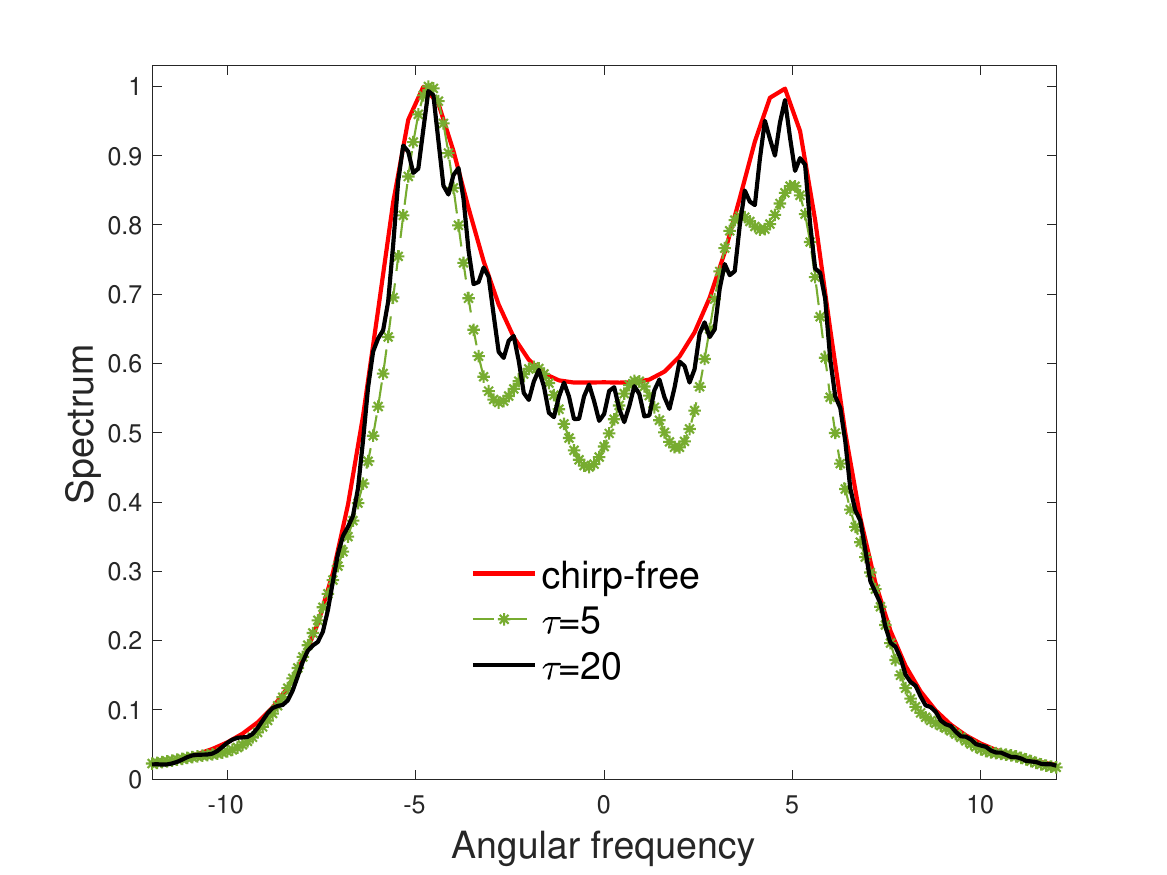}}
	\caption{
     Chirped CRBS spectra for a polyatomic Maxwell gas at different chirp duration $\tau$, with the rotational degrees of freedom $d_r=2$, rotational collision number $Z=1$, translational Eucken factor $f_t=2.4$, and rotational Eucken factor $f_r=1.5$. 
	}
	\label{fig:chirp_Maxwell_polyatomic}
\end{figure}

Figure~\ref{fig:chirp_Maxwell_polyatomic} shows chirped CRBS spectra of a polyatomic Maxwell gas for different chirp durations. Similar to the monatomic case, a short chirp duration ($\tau = 5$) produces asymmetric spectra and pronounced ripples in the Rayleigh peak. As the chirp duration increases, both the degree of spectral asymmetry and the ripple amplitude decrease, while the ripple frequency increases.

\section{Conclusions}

In this work, we developed a fast deterministic solver based on the linearized Boltzmann equation to compute CRBS spectra with high accuracy and efficiency. Each line shape can be obtained within about one minute, making systematic parametric studies feasible.
Our analysis revealed that the CRBS spectrum is highly sensitive to the intermolecular potential in the kinetic regime, where the magnitude of the Rayleigh peak depends strongly on the viscosity index. In particular, when the Knudsen number is fixed, gases modeled by softer potentials exhibit lower Rayleigh peaks, whereas hard-sphere gases yield enhanced Rayleigh scattering compared with Maxwellian molecules. These trends were interpreted in terms of the velocity dependence of the collision frequency and the resulting effective Knudsen number. We further found that increasing the chirp rate produces asymmetric line shapes and fine oscillatory structures near the Rayleigh peak. 



The present computational framework provides both physical insights and a practical tool for interpreting experimental data, paving the way for high-fidelity non-intrusive diagnostics of rarefied gas flows.

\bibliographystyle{elsarticle-num}
\bibliography{thesisBib}

\begin{thebibliography}{10}
\expandafter\ifx\csname url\endcsname\relax
  \def\url#1{\texttt{#1}}\fi
\expandafter\ifx\csname urlprefix\endcsname\relax\def\urlprefix{URL }\fi
\expandafter\ifx\csname href\endcsname\relax
  \def\href#1#2{#2} \def\path#1{#1}\fi

\bibitem{Tenti1974}
G.~Tenti, C.~Boley, R.~Desai, {On the kinetic model description of
  Rayleigh-Brillouin scattering from molecular gases}, Can. J. Phys. 52 (1974)
  285.

\bibitem{srbs_scale}
V.~Ghaem-Maghami, A.~D. May, {Rayleigh-Brillouin spectrum of compressed He, Ne
  and Ar. I. Scaling}, Phys. Rev. A 22 (1980) 692.

\bibitem{Witschas2014}
B.~Witschas, C.~Lemmerz, O.~Reitebuch, {Daytime measurements of atmospheric
  temperature profiles 2--15km by lidar utilizing Rayleigh--Brillouin
  scattering}, Opt. Lett. 39 (2014) 1972--1975.

\bibitem{Gu2014OL}
Z.~Gu, W.~Ubachs, W.~van~de Water, {Rayleigh-Brillouin scattering of carbon
  dioxide}, Opt. Lett. 39 (2014) 3301.

\bibitem{GuRBS2015}
Z.~Gu, W.~Ubachs, W.~Marques, W.~van~de Water, {Rayleigh-Brillouin} scattering
  in binary-gas mixtures, Phys. Rev. Lett. 114 (2015) 243902.

\bibitem{Pan2002}
X.~Pan, M.~N. Shneider, R.~B. Miles, {Coherent Rayleigh-Brillouin Scattering},
  Phys. Rev. Lett. 89~(18) (2002) 183001.

\bibitem{CRBS_JCP}
A.~S. Meijer, A.~S. de~Wijn, M.~F.~E. Peters, N.~J. Dam, W.~van~de Water,
  Coherent {Rayleigh-Brillouin} scattering measurements of bulk viscosity of
  polar and nonpolar gases, and kinetic theory, J. Chem. Phys. 133 (2010)
  164315.

\bibitem{Cornella}
B.~M. Cornella, S.~F. Gimelshein, M.~N. Shneider, T.~C. Lilly, A.~D. Ketsdever,
  {Experimental and numerical analysis of narrowband coherent
  Rayleigh--Brillouin scattering in atomic and molecular species}, Opt. Exp.
  20~(12) (2012) 12975.

\bibitem{Shneider2013}
M.~N. Shneider, S.~F. Gimelshein, {Application of coherent Rayleigh-Brillouin
  scattering for in situ nanoparticle and large molecule detection}, Appl.
  Phys. Lett. 102 (2013) 173109.

\bibitem{barker2013}
A.~Gerakis, M.~N. Shneider, P.~F. Barker, {Single-shot coherent
  Rayleigh--Brillouin scattering using a chirped optical lattice}, Opt. Lett.
  38~(21) (2013) 4449.

\bibitem{Suzuki2024PRA}
S.~Suzuki, A.~Gerakis, K.~Hara, Effects of the chirp rate on single-shot
  coherent {Rayleigh-Brillouin} scattering, Phys. Rev. A 110 (2024) 033519.

\bibitem{Bruno2019CPL}
D.~Bruno, A.~Frezzotti, Dense gas effects in the {Rayleigh-Brillouin scattering
  spectra of SF6}, Chem. Phys. Lett. 731 (2019) 136595.

\bibitem{Bird1994}
G.~A. Bird, Molecular Gas Dynamics and the Direct Simulation of Gas Flows,
  Oxford Science Publications, Oxford University Press Inc, New York, 1994.

\bibitem{Wu-2022}
L.~Wu, Rarefied Gas Dynamics: Kinetic Modeling and Multi-Scale Simulation,
  Springer, 2022.

\bibitem{Lei_AIP}
L.~Wu, C.~White, T.~J. Scanlon, J.~M. Reese, Y.~Zhang, Coherent
  {Rayleigh-Brillouin} scattering: Influence of the intermolecular potential,
  {AIP Conf. Proc.} 1628 (2014) 648.

\bibitem{Lei2013}
L.~Wu, C.~White, T.~J. Scanlon, J.~M. Reese, Y.~H. Zhang, {Deterministic
  numerical solutions of the {B}oltzmann equation using the fast spectral
  method}, J. Comput. Phys. 250 (2013) 27--52.

\bibitem{LeiJFM2015}
L.~Wu, C.~White, T.~J. Scanlon, J.~M. Reese, Y.~H. Zhang, A kinetic model of
  the {Boltzmann equation for} non-vibrating polyatomic gases, J. Fluid Mech.
  763 (2015) 24--50.

\bibitem{CE}
S.~Chapman, T.~Cowling, {The Mathematical Theory of Non-uniform Gases},
  Cambridge University Press, 1970.

\bibitem{wuPoF2015}
L.~Wu, H.~H. Liu, Y.~H. Zhang, J.~M. Reese, {Influence of intermolecular
  potentials on rarefied gas flows: Fast spectral solutions of the Boltzmann
  equation}, Phys. Fluids 27 (2015) 082002.

\bibitem{GrossJackson1959}
E.~P. Gross, E.~A. Jackson, Kinetic models and the linearized {B}oltzmann
  equation, Phys. Fluids 2~(4) (1959) 432--441.

\bibitem{Holway1966}
L.~H. Holway, New statistical models for kinetic theory: methods of
  construction, Phys. Fluids 9 (1966) 1658--1673.

\bibitem{Shakhov_S}
E.~M. Shakhov, {Generalization of the Krook kinetic relaxation equation}, Fluid
  Dyn. 3~(5) (1968) 95--96.

\bibitem{Pan2005PRA}
X.~Pan, M.~N. Shneider, R.~B. Miles, {Power spectrum of coherent
  Rayleigh-Brillouin scattering in carbon dioxide}, Phys. Rev. A 71 (2005)
  045801.

\bibitem{wangchang_uhlenbeck_1952}
C.~S. Wang~Chang, G.~E. Uhlenbeck, On the propagation of sound in monatomic
  gases, Report M999, University of Michigan, Engineering Research Institute
  (1952).

\bibitem{Pan2004}
X.~Pan, M.~N. Shneider, R.~B. Miles, {Coherent Rayleigh-Brillouin scattering in
  molecular gases}, Phys. Rev. A 69 (2004) 033814.

\bibitem{WangCS}
C.~S. Wang-Chang, G.~E. Uhlenbeck, {Transport Phenomena in Polyatomic Gases},
  University of Michigan Engineering Research Rept. No. CM-681, 1951.

\bibitem{Wu2020JFM}
L.~Wu, Q.~Li, H.~Liu, W.~Ubachs, Extraction of the translational {Eucken}
  factor from light scattering by molecular gas, J. Fluid Mech. 901 (2020) A23.

\end{thebibliography}
\end{document}